\documentclass{ws-procs9x6}
\def\be{\begin{equation}}
\def\ee{\end{equation}}
\def\ba{\begin{eqnarray}}
\def\ea{\end{eqnarray}}
\def\OL0{\Omega_{\Lambda}^{(0)}}

\def\Om0{\Omega_{m}^{(0)}}
\def\Or0{\Omega_{r}^{(0)}}

\def\fr{\frac}

\begin{document}

\title{Coupled Quintessence and CMB 
}

\author{Seokcheon Lee 
}

\address{Institute of Physics,
Academia Sinica, \\ Taipei, Taiwan 11529, R.O.C.\\
E-mail: skylee@phys.sinica.edu.tw}

\maketitle

\abstracts{We revise the stability of the tracking solutions and
briefly review the potentials of quintessence models. We discuss
the evolution of linear perturbations for $V(\phi) = V_{0}
\exp(\lambda \phi^2/2)$ potential in which the scalar field is
non-minimally coupled to cold dark matter. We consider the effects
of this coupling on both cosmic microwave background temperature
anisotropies and matter perturbations. We find that the
phenomenology of this model is consistent with current
observations up to the coupling power $n_{c} \leq 0.01$ while
adopting the current parameters measured by WMAP,
$\Omega_{\phi}^{(0)}=0.76$, $\Omega_{cdm}^{(0)}=0.191$,
$\Omega_{b}^{(0)}=0.049$, and $h=0.70$. Upcoming cosmic microwave
background observations continuing to focus on resolving the
higher peaks may put strong constraints on the strength of the
coupling.}

\section{Introduction}
If we treat Type Ia supernovae (SNe Ia) as standardized candles,
then the Hubble diagram of them shows that the expansion of the
Universe is currently accelerating \cite{SCP}. Combining
measurements of the acoustic peaks in the angular power spectrum
of the cosmic microwave background (CMB) anisotropy \cite{CMB} and
the matter power spectrum of large scale structure (LSS) which is
inferred from galaxy redshift surveys like the Sloan Digital Sky
Survey (SDSS) \cite{SDSS} and the $2$-degree Field Galaxy Redshift
Survey ($2$dFGRS) \cite{2dFGRS} has also confirmed that a
component with negative pressure (dark energy) should be added to
the matter component to make up the critical density today.

A quintessence field is a dynamical scalar field leading to a time
dependent equation of state (EOS), $\omega_{\phi}$. The
possibility that a scalar field at early cosmological times
follows an attractor-type solution and tracks the evolution of the
visible matter-energy density has been explored \cite{Ratra}. This
may help alleviate the severe fine-tuning associated with the
cosmological constant problem. However we need to investigate the
tracking condition and its stability at the matter dominated epoch
carefully \cite{Lee1}.

Are there experimental ways of checking for the existence or
absence of dark energy in the form of quintessence? There are
several different observational effects of matter coupling to the
scalar field on CMB spectra and matter power spectrum compared to
the minimally coupled models \cite{LLN}. And this can be used to
check the existence of quintessence.

This paper is organized as follows. We briefly investigate the
condition and the stability of tracker solution and review the
potentials of quintessence models in the next section. In Sec.
III, we show the coupling effects on CMB and matter power
spectrum. We conclude in the last section.

\section{Quintessence Models and Tacker Solutions}
Many models of quintessence have a tracker behavior, which solves
the ``coincidence problem'' ({\it i.e.} initial condition). In
these models, the quintessence field has a density which closely
tracks (but is less than) the radiation density until
matter-radiation equality, which triggers quintessence to start
behaving as dark energy, eventually dominating the Universe. This
naturally sets the low scale of the dark energy. However the
present small value of dark energy density still cannot be solved
with quintessence (``fine-tuning problem'').

Since the energy density of the scalar field generally decreases
more slowly than the matter energy density, it appears that the
ratio of the two densities must be set to a special, infinitesimal
value in the early Universe in order to have the two densities
nearly coincide today. To avoid this initial conditions problem we
focus on tracker fields.

\subsection{Tracking Condition}\label{subsec:TC}

We rely details of this section on the reference \cite{Lee1} due
to the shortage of space. We introduce new quantity $\theta$ which
is related to the ratio of the kinetic energy and the potential
energy of the scalar field : \be 2 \theta = \ln \frac{KE}{PE} =
\ln \frac{1 + \omega_{\phi}}{1 - \omega_{\phi}} \label{theta} \ee
$\theta$ can have any value and especially positive $\theta$ means
the kinetic energy dominated era and negative $\theta$ indicates
the potential energy dominated one. Now we can define EOS as the
function of this new quantity $\theta$ : \be \omega_{\phi} = \tanh
\theta \label{omegaphi} \ee The ``tracker equation'' ($\Gamma
\equiv V''V/(V')^2$) can be expressed by $\theta$ ; \be \Gamma = 1
+ \frac{3}{2} \frac{(A \omega_{r} -
\omega_{\phi})}{(1+\omega_{\phi})}\frac{(1 - \Omega_{\phi})}{(3 +
\tilde{\theta})} - \frac{1 - \omega_{\phi}}{2(1 + \omega_{\phi})}
\frac{\tilde{\theta}}{(3 + \tilde{\theta})} -\frac{1}{(1 +
\omega_{\phi})} \frac{\stackrel{\approx}{\theta}}{(3 +
\tilde{\theta})^2}\label{Gamma} \ee where $\omega_{r}$ is EOS of
the radiation, $A=a_{eq}/(a+a_{eq})$, $a_{eq}$ is the scale factor
when the energy density of the radiation and that of the matter
become equal, and tilde means the derivative with respect to $x
=\ln a = - \ln(1+z)$ \footnote{Where we put the present value of
scale factor, $a^{(0)}$ as one.}. This equation looks like quite
different from the well known tracker equation. But when we choose
the early Universe constraints ($A \simeq 1$ and $\Omega_{\phi}
\simeq 0$) we can get the well known tracker equation. \ba \Gamma
&\simeq& 1 + \frac{3}{2} \frac{(\omega_{r} -
\omega_{\phi})}{(1+\omega_{\phi})}\frac{1}{(3 + \tilde{\theta})} -
\frac{1 - \omega_{\phi}}{2(1 + \omega_{\phi})}
\frac{\tilde{\theta}}{(3 + \tilde{\theta})} -\frac{1}{(1 +
\omega_{\phi})} \frac{\stackrel{\approx}{\theta}}{(3 +
\tilde{\theta})^2} \nonumber \\
&=& 1 + \frac{(\omega_{r} - \omega_{\phi})}{2(1+\omega_{\phi})} -
\frac{1 + \omega_{r} - 2\omega_{\phi}}{2(1 + \omega_{\phi})}
\frac{\tilde{\theta}}{(3 + \tilde{\theta})} -\frac{1}{(1 +
\omega_{\phi})} \frac{\stackrel{\approx}{\theta}}{(3 +
\tilde{\theta})^2} \label{GammaS} \ea When we can ignore the
change of $\theta$ ($i.e.$ when $\omega_{\phi}$ is almost
constant), we can get the tracking solution. From the above
tracking equation we can check this condition. \be \Gamma \simeq 1
+ \frac{1}{2} \frac{(A \omega_{r} - \omega_{\phi})}{1 +
\omega_{\phi}} (1 - \Omega_{\phi}) \label{Gamma1} \ee This
equation can be rearranged to see the behavior of the EOS as
following. \be \omega_{\phi} \simeq \frac{\omega_{r} A (1 -
\Omega_{\phi}) - 2 ( \Gamma - 1 )}{(1 - \Omega_{\phi}) + 2 (
\Gamma - 1)} \label{omega1} \ee To investigate this equation more
carefully, we define the new quantities.
\ba Q &=& (1- \Omega_{\phi}) \label{Q} \\
F &=& (\Gamma - 1) \label{F1} \ea where $Q$ shows the energy
information of the Universe and $F$ depends on the form of the
given potential. 
With these we can rewrite the equation (\ref{omega1}). \be
\omega_{\phi} \simeq \omega_{r} \frac{Q}{Q+2F} A - \frac{2F}{Q+2F}
\label{omega2} \ee In the reference \cite{Ratra}, this equation is
expressed as : \be \omega_{\phi} \simeq \frac{\omega_{r} -
2F}{1+2F} \label{omega2S} \ee This equation (\ref{omega2S}) can be
true only when $Q, A \simeq 1$, which can be satisfied at the
early Universe and not at the late one. So with this equation, it
is not proper to check the evolution of tracking solutions at late
Universe. Instead we should use the equation (\ref{omega2}) to
check the evolution of the tracking solutions. Before checking the
properties of this equation, we should notice that $Q$ is always
positive and has the interval as $0 \leq Q \leq 1$. $F$ can be
positive or negative based on the given shape of the potential.

\subsection{Stability Of Tracker Solution}\label{SoTS}

We need to check that solutions with $\omega_{\phi}$,which is not
equal to the tracker solution value ($\omega_{0}$) can converge to
the tracker ones ($i.e.$ Are tracker solutions stable?). To check
this we need to check the small deviation ($\delta \omega$) of the
tracker solution of EOS. \be \omega_{\phi} = \omega_{0} + \delta
\omega \label{deltaomega} \ee If we insert this into the tracker
equation (\ref{Gamma}), then we have following. \be
\stackrel{\approx}{\delta \omega} + \frac{3}{2} \Bigl[ (A
\omega_{r} - \omega_{0})(1 - \Omega_{\phi}) + ( 1 - \omega_{0})
\Bigr] \tilde{\delta \omega} + \frac{9}{2} (1 - \omega_{0})
\Bigl[(1 + A \omega_{r})(1 - \Omega_{\phi}) \Bigr] \delta \omega =
0 \label{deltaomega1} \ee where we use the tracking condition
(\ref{Gamma1}). The general solution to this nonlinear
differential equation cannot be obtained analytically. But this
equation can be simplified as follow in the early Universe
constraints. \be \stackrel{\approx}{\delta \omega} + \frac{3}{2}
\Bigl[ (1 + \omega_{r}) - 2 \omega_{0} \Bigr] \tilde{\delta
\omega} + \frac{9}{2} (1 - \omega_{0})(1 + \omega_{r}) \delta
\omega \simeq 0 \label{deltaomega2} \ee The solution of this
equation is \be \delta \omega \propto a^{\gamma_{1}}
\label{deltaomega21} \ee where \be \gamma_{1} = -\frac{3}{2}
\Bigl[ \frac{1}{2}(1 + \omega_{r}) - \omega_{0} \Bigr] \pm
\frac{i}{2} \sqrt{18(1 + \omega_{r})(1 - \omega_{0}) - 9
\Bigr[\frac{1}{2}(1 + \omega_{r}) - \omega_{0} \Bigr]^2}
\label{gamma1} \ee The real part of this is negative for
$\omega_{0}$ less than $2/3$. So $\delta \omega$ will decays
exponentially and solution reaches to the tracking one. In
addition to this it also oscillates due to the second term.  For
the late Universe case, we can change this equation. In the late
Universe $A$ goes to zero and $\Omega_{\phi}$ is not zero. With
these we can modify the general equation (\ref{deltaomega1}). \be
\stackrel{\approx}{\delta \omega} + \frac{3}{2} \Bigl[ (1 +
\Omega_{\phi}\omega_{0}) - 2 \omega_{0} \Bigr] \tilde{\delta
\omega} + \frac{9}{2} (1 - \omega_{0})(1 - \Omega_{\phi}) \delta
\omega \simeq 0 \label{deltaomega3} \ee We can repeat the similar
step to find the solution of this equation if we assume that
$\Omega_{\phi}$ is almost constant. \be \delta \omega \propto
a^{\gamma_{2}} \label{deltaomega31} \ee where  \be \gamma_{2} =
-\frac{3}{2} \Bigl[ \frac{1}{2}(1 + \Omega_{\phi}\omega_{0}) -
\omega_{0} \Bigr] \pm \frac{i}{2} \sqrt{18(1 + \Omega_{\phi})(1 -
\omega_{0}) - 9 \Bigr[\frac{1}{2}(1 + \Omega_{\phi}\omega_{0}) -
\omega_{0} \Bigr]^2} \label{gamma2} \ee The real part of this
solution can be negative if $\omega_{0}$ satisfies following. \be
\omega_{0} < \frac{1}{(2 - \Omega_{\phi})} \label{gamma21} \ee
where $1 \leq (2 - \Omega_{\phi}) \leq 2$ for the entire history
of Universe.

\subsection{Quintessence Potentials}\label{QV}

We display the potentials of the quintessence models in Table
\ref{table1}. Any detail of each model can be found in each
reference.

\begin{table}[t]
\tbl{Quintessence models.} {\footnotesize
\begin{tabular}{@{}ccc@{}}
\hline
{} &{} &{} \\[-1.5ex]
Potential & Reference & Properties \\[1ex]
\hline
{} &{} &{} \\[-1.5ex]
$V_0\exp{(-\lambda\phi)}$ & Ratra \& Peebles (1988), Wetterich
(1988) \cite{Ratra}
& $\omega = \lambda^2/3 -1$\\[1ex]
& Ferreira \& Joyce (1998) \cite{Ratra} & $ \lambda > 5.5-4.5,
\Omega < 0.1-0.15$
\\[1ex]
$V_0/\phi^\alpha, \alpha > 0$ &  Ratra \& Peebles (1988)
\cite{Ratra} & $
\omega > -0.7$\\[1ex]
$m^2\phi^2, \lambda\phi^4$ &  Frieman et al (1995) \cite{Fri} &
PNGB, $M^4[\cos(\phi/f)+1]$\\[1ex]
$V_0(\exp{M_p/\phi} - 1)$ & Zlatev, Wang \& Steinhardt (1999)
\cite{Zal} &
$\Omega_{m} \geq 0.2, \omega < -0.8$\\[1ex]
$V_0\exp{(\lambda\phi^2)}/\phi^\alpha $ & Brax \& Martin
(1999,2000) \cite{Brax} & $\alpha \geq 11
, \omega \simeq -0.82$\\[1ex]
$V_0(\cosh{\lambda\phi} - 1)^p$ & Sahni \& Wang (2000) \cite{SaWa}
& $p <1/2, \omega < -1/3$\\[1ex]
$V_0 \sinh^{-\alpha}{(\lambda\phi)}$ & Sahni \& Starobinsky
(2000) \cite{SaSt}, & early time : inverse power \\[1ex]
& Ure\~{n}a-L\'{o}pez \& Matos (2000) \cite{Urena} & late time : exponential\\[1ex]
$V_0(e^{\alpha\kappa\phi} + e^{\beta\kappa\phi})$ & Barreiro,
Copeland \& Nunes (2000) \cite{Bar} & $\alpha > 5.5, \beta <0.8, \omega < -0.8$ \\[1ex]
$V_0[(\phi - B)^\alpha + A]e^{-\lambda\phi}$ & Albrecht \& Skordis
(2000) \cite{Alb} & $\omega \sim -1$
\\[1ex]
$V_0 \exp[{\lambda (\phi/M_p)^2}]$ &Lee, Olive, \& Pospelov (2004)
\cite{LOP} & $\omega \sim -1$
\\[1ex]
$V_0 \cosh[{\lambda \phi/M_p}]$ & Lee, Olive, \& Pospelov (2004) \cite{LOP}
& $\omega \sim -1$\\[1ex]
\hline
\end{tabular}\label{table1} }
\vspace*{-13pt}
\end{table}


\section{Coupled Quintessence}

The general equation for the interaction of a light scalar field
$\varphi$ with matter is,
\begin{eqnarray}
S_{\phi} &=& \int d^4x \sqrt{-g} \Biggl\{ \frac{\bar{M}^2}{2}
[\partial^{\mu} \phi
\partial_{\mu} \phi - R] - V(\phi)
- \frac{B_{Fi}(\phi)}{4}F^{(i)}_{\mu\nu}F^{(i)\mu\nu} \nonumber
\\ && + \sum_j [\bar\psi_j iD\!\!\!\!/ \psi_j -
B_j(\phi)m_j\bar\psi_j\psi_j]\Biggr\}. \label{lagrangian}
\end{eqnarray}
The coupling gives rise to the additional mass and source terms of
the evolution equations for CDM and scalar field perturbations.
This also affects the perturbation of radiation indirectly through
the background bulk and the metric perturbations
\cite{LLN},\cite{Bean2}. The value of the energy density contrast
of the CDM ($\Omega_{c}$) is increased in the past when the
coupling is increased. We specify the potential and the coupling
as in the reference \cite{LLN},\cite{LOP}. \be V(\phi) = V_{0}
\exp \Bigl( \frac{\lambda\phi ^2 }{2} \Bigr),  \hspace{0.5in} \exp
[B_{c}(\phi)] = \Biggl(\frac{b_c+V(\phi)/V_0}{1+b_c}\Biggr)^{n_c}
 \label{VBF} \ee

\subsection{CMB}
\begin{figure}[ht]
\centerline{\epsfxsize=2.5in \epsfbox{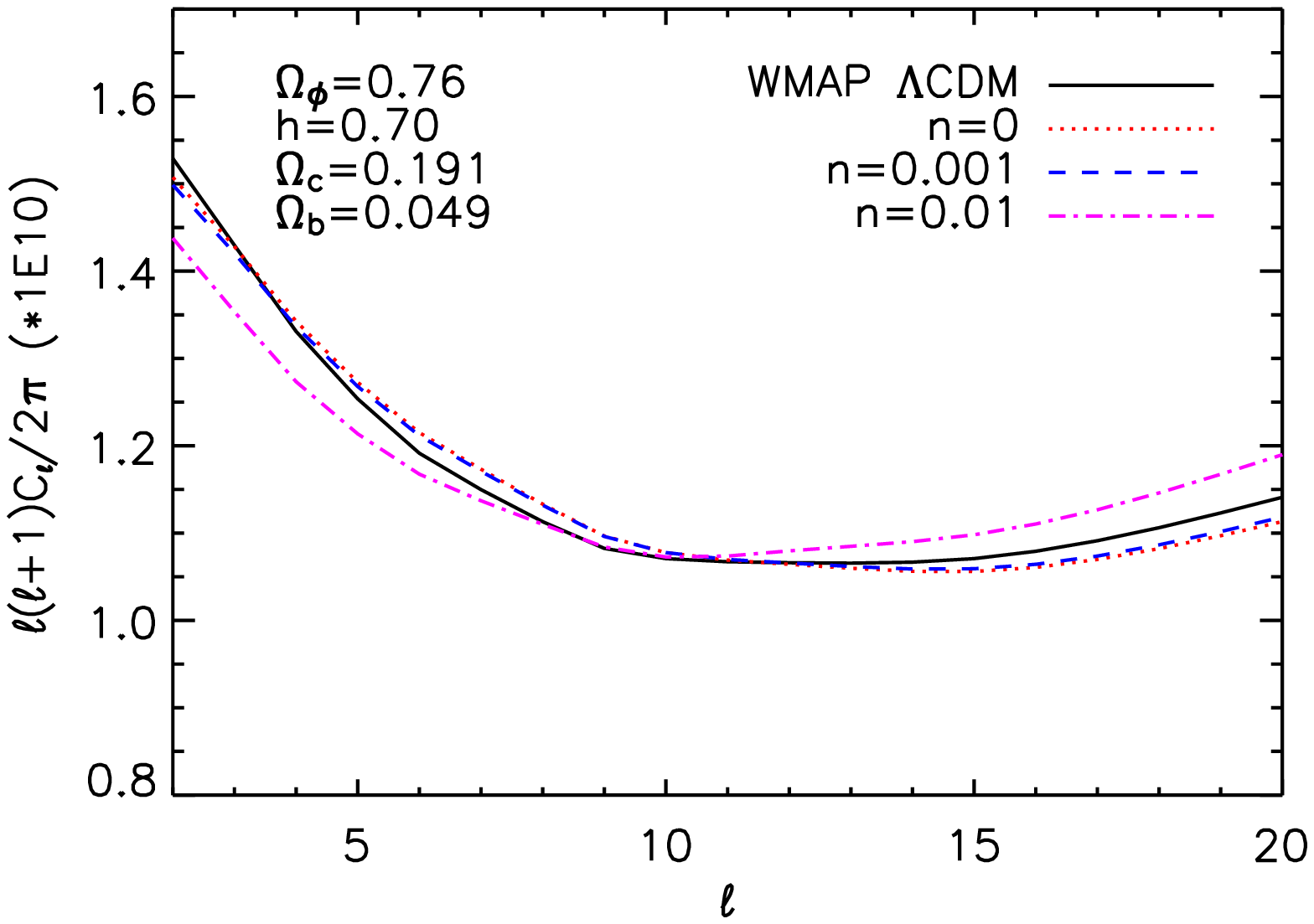}
\epsfxsize=2.5in \epsfbox{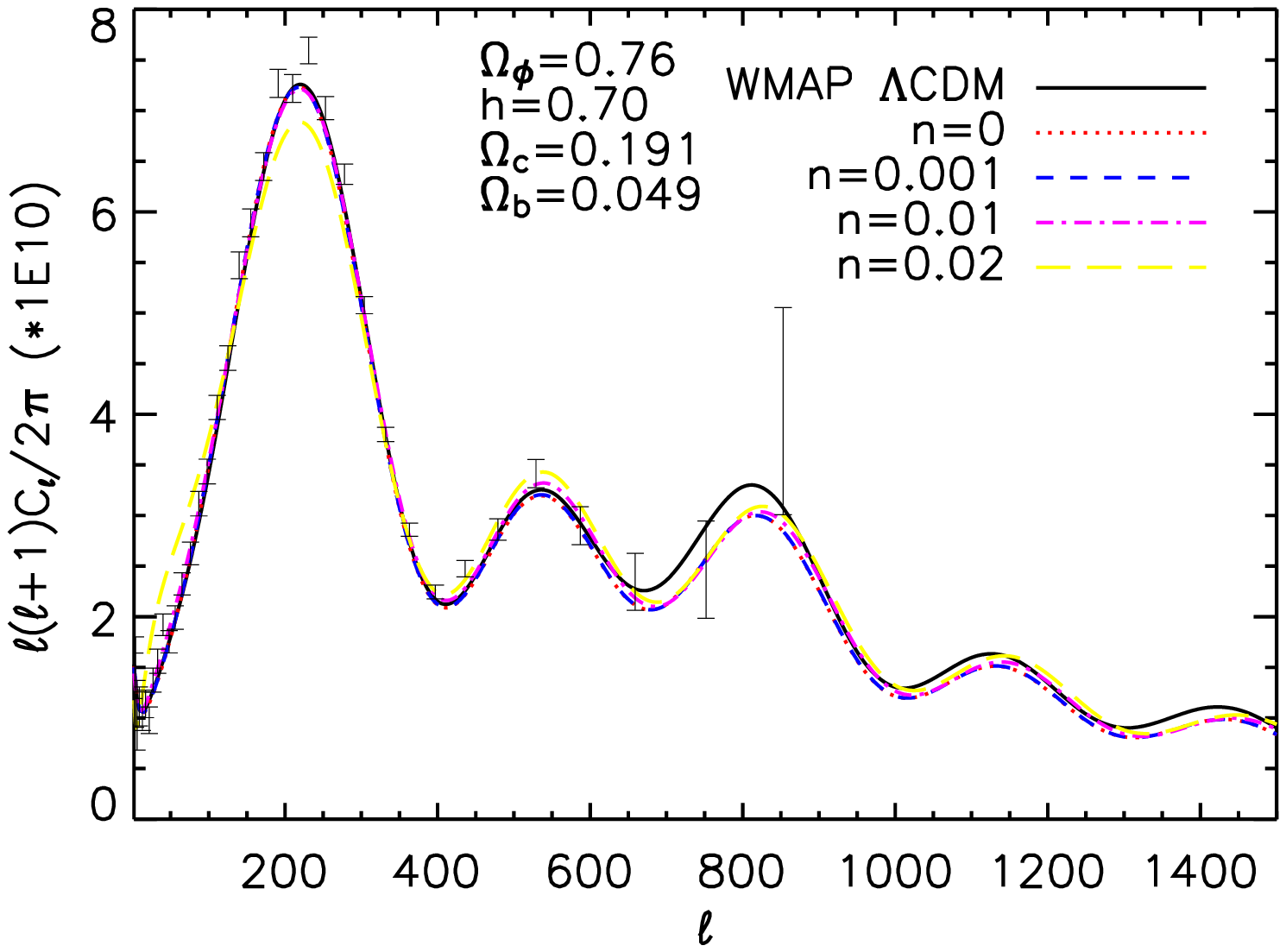} }
\caption{ (a) CMB large-scale anisotropy power spectra of
$\Lambda$CDM (solid line), minimally coupled $n_{c} = 0$ (dotted
line), and non-minimally coupled $n_{c} = 10^{-3}, 10^{-2}$
(dashed, dash-dotted line respectively) quintessence models. (b)
Same spectra for the entire scales. \label{fig:cl}}
\end{figure}

Now, we investigate the effects of non-minimal coupling of a
scalar field to the CDM on the CMB power spectrum. Firstly, the
Newtonian potential at late times changes more rapidly as the
coupling increases. This leads to an enhanced ISW effect. Thus we
have a relatively larger $C_{\ell}$ at large scales ({\it i.e.}
small $\ell$). Thus, if the CMB power spectrum normalized by COBE,
then we will have smaller quadrupole \cite{COBE}. This is shown in
the first panel of Figure~\ref{fig:cl}. One thing that should be
emphasized is that we use different parameters for the
$\Lambda$CDM and the coupled quintessence models to match the
amplitude of the first CMB anisotropy peak. The parameter used for
the quintessence model is indicated in Figure~\ref{fig:cl} ({\it
i.e.} $\Omega_{\phi}^{(0)} = 0.76$, $\Omega_{m}^{(0)} = 0.191$,
$\Omega_{b}^{(0)} = 0.049$, and $h = 0.7$, where $h$ is the
present Hubble parameter in the unit of $100 {\rm km s^{-1}
Mpc^{-1}}$). However, these parameters are well inside the $1 \,
\sigma$ region given by the WMAP data. We use the WMAP parameters
for the $\Lambda$CDM model ({\it i.e.} $\Omega_{\phi}^{(0)} =
0.73$, $\Omega_{m}^{(0)} = 0.23$, $\Omega_{b}^{(0)} = 0.04$, and
$h = 0.72$) \footnote{Our data prefers WMAP 3 year data to WMAP 1
year one.}. In both models we use the same spectral index $n_{s}
=1$. The heights of the acoustic peaks at small scales ({\it i.e}
large $\ell$) can be affected by the following two factors. One is
the fact that the scaling of the CDM energy density deviates from
that of the baryon energy density. Therefore for the given CDM and
baryon energy densities today, the energy density contrast of
baryons at decoupling ($\Omega_{b}^{(ls)}$) is getting lower as
the coupling is being increased. This suppresses the amplitude of
compressional (odd number) peaks while enhancing rarefaction (even
number) peaks. The other is that for models normalized by COBE,
which approximately fixes the spectrum at $\ell \simeq 10$, the
angular amplitude at small scales is suppressed in the coupled
quintessence. This is shown in the second panel of
Figure~\ref{fig:cl}. The third peak in this model is smaller than
that in the $\Lambda$CDM model.

\subsection{Matter Power Spectrum}

\begin{figure}[ht]
\centerline{\epsfxsize=4.1in
\epsfbox{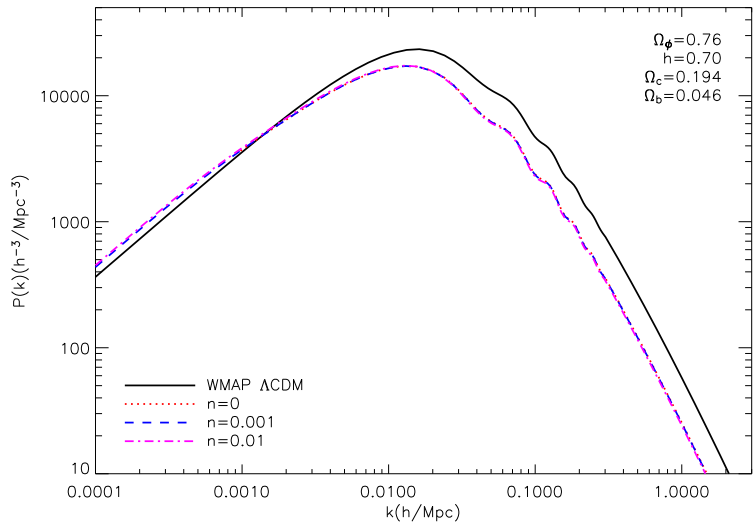}} \caption{Matter power
spectra for the models using the same parameters in
Figure~\ref{fig:cl}. \label{fig:mp}}
\end{figure}

The coupling of quintessence to the CDM can change the shape of
matter power spectrum because the location of the turnover
corresponds to the scale that entered the Hubble radius when the
Universe became matter-dominated. This shift on the scale of
matter and radiation equality is indicated \be a_{eq} \simeq
\fr{\rho_{r}^{(0)}}{\rho_{c}^{(0)}} \exp [B_{c}(\phi_{0}) -
B_{c}(\phi_{eq})], \label{aeq} \ee where $\rho_{r}^{(0)}$ and
$\rho_{c}^{(0)}$ are the present values of the energy densities of
radiation and CDM respectively, and the approximation comes from
the fact that the present energy density of CDM is bigger than
that of baryons ($\rho_{c}^{(0)} > \rho_{b}^{(0)}$). This is
indicated in Figure~\ref{fig:mp}. Increasing the coupling shifts
the epoch of matter-radiation equality further from the present,
thereby moving the turnover in the power spectrum to smaller
scale. If we define $k_{eq}$ as the wavenumber of the mode which
enters the horizon at radiation-matter equality, then we will
obtain \be k_{eq} = \fr{2 \pi}{\eta_{eq}}. \label{keq} \ee

\section{Conclusions}
We have investigated the tracking condition of the quintessence
models and their stability. We have shown that it is necessary to
distinguish the tracking condition at the matter dominated epoch
and at the radiation dominated one.

We have considered the CMB anisotropy spectrum and the matter
power spectrum for the non-minimally coupled models. Additional
mass and source terms in the Boltzmann equations induced by the
coupling give the rapid changes of the Newtonian potential $\Phi$
and enhance the ISW effect in the CMB power spectrum. The
modification of the evolution of the CDM, $\rho_{c} =
\rho_{c}^{(0)} a^{-3 + \xi}$, changes the energy density contrast
of the CDM at early epoch. We have adopted the current
cosmological parameters measured by WMAP within $1 \sigma$ level.
With the COBE normalization and the WMAP data we have found the
constraint of the coupling $n_{c} \leq 0.01$. The locations and
the heights of the CMB anisotropy peaks have been changed due to
the coupling. Especially, there is a significant difference for
the heights of  the second and the third peaks among the models.
Thus upcoming observations continuing to focus on resolving the
higher peaks may constrain the strength of the coupling. The
suppression of the amplitudes of the matter power spectra could be
lifted by a bias factor. However, a detailed fitting is beyond the
scope of this paper. The turnover scale of the matter power
spectrum may be also used to constrain the strength of the
coupling $n_{c}$.

\section*{Acknowledgments}
We thank ICGA7 organizing committee, J. M. Nester, C-M. Chen, and
J-P. Hsu for their hospitality and for organizing such a nice
meeting.


\begin{thebibliography}{0}
\bibitem{SCP} A.~G.~Riess {\it et al.}  [Supernova Search Team Collaboration],
Astron.\ J.\  {\bf 116}, 1009 (1998) [astro-ph/9805201];
Astrophys.\ J.\ {\bf 607}, 665 (2004) [astro-ph/0402512];
S.~Perlmutter {\it et al.}  [Supernova Cosmology Project
Collaboration],
Astrophys.\ J.\  {\bf 517}, 565 (1999) [astro-ph/9812133];
N.~A.~Bahcall, J.~P.~Ostriker, S.~Perlmutter, and
P.~J.~Steinhardt,
Science {\bf 284}, 1481 (1999) [astro-ph/9906463]; T.~Padmanabhan
and T.~R.~Choudhury, Mon.\ Not.\ Roy.\ Astron.\ Soc.\ {\bf 344},
823 (2003) [astro-ph/0212573]; Astron.\ Astrophys.\ {\bf 429}, 807
(2005) [astro-ph/0311622]; P.~Astier {\it et al.},
[astro-ph/0510447].

\bibitem{CMB} C. L. Bennett {\it et al.}, Astrophys.\ J.\ Suppl
. Ser {\bf 148}, 1 (2003) [astro-ph/0302207]; D. N.~Spergel {\it
et al.}, Astrophys.\ J.\ Suppl. Ser. {\bf 148}, 175 (2003)
[astro-ph/0302209].

\bibitem{SDSS} M.~Tegmark {\it et al.}, Phys.\ Rev.\ D ~{\bf 69},
103501 (2004) [astro-ph/0310723].

\bibitem{2dFGRS} M.~Colless {\it et al.}, Mon. Not. Roy. Astron.
Soc. {\bf 328}, 1039 (2001) [astro-ph/0106498];
[astro-ph/0306581].

\bibitem{Ratra} B.~Ratra and P.~J.~E.~Peebles, Phys.\ Rev.\ D {\bf 37}, 3406 (1988);
Astrophys. J. {\bf 325}, L117 (1988); C.~Wetterich, Nucl.\ Phys.\
B {\bf 302}, 668 (1988); 
P.~G.~Ferreira and M.~Joyce, Phys.\ Rev.\ D {\bf 58}, 023503
(1998) [astro-ph/9711102].

\bibitem{Lee1} S.~Lee, PhD\ Thesis, University of Minnesota,
(2005).

\bibitem{LLN} S.~Lee, G-C.~Liu, and K-W.~Ng, Phys.\ Rev.\ D {\bf73}, 083516 (2006)
[astro-ph/0601333].

\bibitem{Fri} J.~Frieman, C.~Hill, A.~Stebbins, and I.~Waga,
Phys.\ Rev.\ Lett. {\bf75}, 2077 (1995) [astro-ph/9505060].

\bibitem{Zal} I.~Zlatev, L.~Wang, and P.~J.~Steinhardt,
Phys.\ Rev.\ Lett. {\bf82}, 896 (1999) [astro-ph/9807002].

\bibitem{Brax} P.~Brax and J.~Martin,
Phys.\ Lett.\ B {\bf468}, 40 (1999) [astro-ph/9905040] ; Phys.\
Rev.\ D {\bf 61}, 103502 (2000) [astro-ph/9912046].

\bibitem{SaWa} V.~Sahni and L.~Wang,
Phys.\ Rev.\ D {\bf62}, 103517 (2000) [astro-ph/9910097].

\bibitem{SaSt} V.~Sahni and A.~A.~Starobinsky,
Int.\ J.\ Mod.\ Phys.\ D {\bf 9}, 373 (2000) [astro-ph/9904398].

\bibitem{Urena} L.~A.~Urena-Lopez and T.~Matos,
Phys.\ Rev.\ D {\bf62}, 081302 (2000) [astro-ph/0003364].

\bibitem{Bar} T.~Barreiro, E.~J.~Copeland, and N.~J.~Nunes
Phys.\ Rev.\ D {\bf61}, 127301 (2000) [astro-ph/9910214].

\bibitem{Alb} A.~Albrecht and C.~Skordis
Phys.\ Rev.\ Lett. {\bf84}, 2076 (2000) [astro-ph/9908085].

\bibitem{LOP} S.~Lee, K.~A.~Olive, and M.~Pospelov
Phys.\ Rev.\ D {\bf70}, 083503 (2004) [astro-ph/0406039].

\bibitem{Bean2} R.~Bean, Phys.\ Rev.\ D {\bf 64}, 123516 (2001) [astro-ph
/0104464].

\bibitem{COBE} M.~White and E.~F.~Bunn, Astrophys.\ J {\bf 450},
477 (1995); Erratum-ibid. {\bf 477}, 460 (1995)
[astro-ph/9503054].

\end{thebibliography}
\end{document}